\documentclass[psfig,twocolumns,a4paper]{article}
\usepackage{latexsym}
\usepackage[dvips]{color}
\usepackage[dvips]{graphicx}   
\usepackage{amsmath}           
\usepackage{amsfonts}          
\usepackage{amssymb}           

\usepackage[english]{babel}

\newcommand{\acknowledgments}{\section*{Acknowledgments}}

\begin{document}

\begin{center}
  \def\PACS{\par\leavevmode\hbox {\it PACS:89.80.+h, 75.10.Nr} (old ones)}

  \Large
  {Disordered Systems, Spanning Trees and SLE}
  \normalsize

  \author
  {
    Davide Fichera,
  }
  Davide Fichera\\
  {Universit\`a degli Studi di Milano - Dip. di Fisica and INFN,
  \\ via Celoria 16, I-20133 Milano  
  \\ \noindent
  \begin{tabular}{cl}
    Mail address:& \tt{David{}e.Fiche{}ra@mi.in{}fn.it} 
  \end{tabular}
}

\date{\today}
\end{center}

\begin{abstract}
We define a minimization problem for paths on planar graphs that, on the 
honeycomb lattice, is equivalent to the exploration path of the critical 
site percolation and than has the same scaling limit of $\mathrm{SLE}_6$. 
We numerically study this model (testing several SLE properties on other 
lattices and with different boundary conditions) and state it in terms of 
spanning trees. This statement of the problem  allows the definition of a 
random growth process for trees 
on two dimensional graphs such that SLE is recovered as a special choice of 
boundary conditions.
\end{abstract}




\section{Introduction}
\label{sec.intro}
Recently, some efforts have been done to relate minimal paths in 
two dimensional disordered systems and SLE processes\footnote{Schramm-Loewner 
Evolution (SLE) is a stochastic process that 
describes the growth of random curves in simply connected two 
dimensional 
domains; for a review see \cite{BauerBernard} and references therein.}, 
see for example 
the boundary walls in Ising Spin Glasses \cite{Hartmann-Amoruso} 
\cite{BernardLeDoussal}. 

In this draft a minimization problem on paths equivalent to $\mathrm{SLE}_6$ 
is in\-tro\-duced on two dimensional honeycomb lattices. 
 
This model is presented in a more general framework (the one of 
spanning trees) and, in order to understand the origin of conformal 
invariance and Markov property, it is studied in some detail also on 
other lattices and with different boundary conditions.  

Such an approach is interesting not only to investigate the possible 
relation of disordered systems with SLE, but also because it could 
give us a deeper 
insight into the understanding of two dimensional disordered systems. 

\section{The model}
\label{sec:themodel}
We are given a planar  two dimensional lattice $\mathcal{G}$ and a set of real 
weights $\omega(f)$ on its faces (plaquettes). 

Any edge $e \equiv (i,j)$ linking two vertices $i$ and $j$, is adjacent to two 
plaquettes: $f_{i,j}$ and $f_{j,i}$ (unless $e$ is an edge on the border of 
$\mathcal{G}$). For a given set of $\{\omega (f)\}$, and a fixed threshold 
$\theta$, we associate to each edge a weight 

\begin{equation}
\label{eqn:makeweights} 
W_{i,j} = [\omega(Pl_{i,j,1}) -\theta] \cdot [\omega(Pl_{i,j,2}) - \theta]
\end{equation} 

Let $(\mathcal{G},W)$ denote the graph with so defined weights on the edges. 

Given a path $\gamma$ of length $N(\gamma) = |\gamma|$ with endpoints $i_0$ 
and $i_N$ we associate to this path the ordered list (in decreasing order):  
$\vec{W} (\gamma) = \mathrm{sort}(\{W_e\}_{e \in \gamma}) = (W_1 (p), \dots , 
W_N (\gamma)) $

We define the order relation "$<$" among paths as follows: 
$\gamma_1 < \gamma_2$ if 
\begin{itemize}
\item Exists $k$ such that $W_j (\gamma_1)  = W_j (\gamma_2)\; \forall j<k$,\, 
 $W_k (\gamma_1)  < W_k (\gamma_2)$;
\item $N(\gamma_1) < N(\gamma_2)$ and $W_j (\gamma_1)  = W_j (\gamma_2) 
\; \forall j \leq N(\gamma_1)$
\end{itemize}
with this definition, either $\gamma_1 \equiv \gamma_2$, or $\gamma_1 
\lessgtr \gamma_2$, i.e. we have a full order provided $W_e \neq 
W_{e^\prime}$ for $e \neq e^\prime$. 

Notice that for each $\gamma_1 < \gamma_2$ exists a $\beta$ such that 
$\forall \beta^\prime \geq \beta$, $\sum_{e\in \gamma_1} \exp(\beta^\prime W_e)  
<  \sum_{e\in \gamma_2} \exp(\beta^\prime W_e)$, so that the function $f_\beta 
(\gamma):= \sum_{e\in\gamma} \exp(\beta W_e)$, in the large $\beta$ limit, 
 is an addictive  cost function  
($f_\beta (\gamma_1 \cup \gamma_2) = f_\beta (\gamma_1) +  f_\beta (\gamma_2)$) 
which reproduce our order relation.

With abuse of language the word cost will be used also for 
the first entry $W_1 (\gamma)$ of the vector $\vec{W}(\gamma)$.

\subsection{Some Remarks}
First of all notice that the optimal path connecting two vertices is always 
a simple path. Call $p_{i,j}$ the optimal path connecting $i$ to $j$. Simple 
reasonings show that $\forall i,j,k,l$ (also coincident), $p_{i,j} \cup 
p_{k,l}$ cannot contain any cycle(as happens for every addictive cost 
function without negative cost loops). Fur\-ther\-more, for our cost function, 
the 
union $T := \cup_{i,j \in V} p_{i,j}$ of all the optimal paths is a tree. 
This tree, spanning for definition, is also the one which minimizes the global 
cost function $\mathcal{H}(T) = \sum{e\in T} W_e$ on the ensemble of the 
spanning tree; it is the   Minimum Spanning Tree\footnote{
The Minimum Spanning Tree of an arbitrary weighted graph is the loopless 
cover of the graph $\mathcal{G}$ that covers every vertex and minimizes the 
sum of the weights on the edges. 
}  
of $(\mathcal{G}, W)$, as one sees analysing the Prim's Algorithm (cfr 
App.~\ref{appendix:MST})

All of this holds for a generic $(\mathcal{G},W)$, but crucially relies 
on our choice of order relation (and thus of optimality for $\gamma_{i,j}$)

Summarizing: given a planar lattice with arbitrary weights on the plaquettes 
we introduced some weights on the edges of the graph as to obtain a graph 
with weighted edges $(\mathcal{G},W)$, a cost function associated to each path 
and an order relation associated to the set of all the paths. 
We stress the fact that the union of all the optimal paths is the MST 
for $(\mathcal{G},W)$. Now we can specialize to a set of planar graphs 
(we will consider only rectangular domains) and to a probability measure 
for the weights on the plaquettes (we will only consider i.i.d. weights 
on the plaquettes).

Consider now a simply connected 
two dimensional domain (e.g. a square) covered with a honeycomb lattice. 
Extract the weights $\omega(\mathcal{G})$ from the dis\-tri\-bu\-tion 
$\chi_{[0,1]}$ and let the threshold be $\theta = 0.5$. 
Fix two different edges on the boundary: $s$ (start) and $t$ (end). 
Constrain all 
the boundary plaquettes on the right of $s$ and $t$ to 
have a cost larger than the threshold and all the other boundary plaquettes 
to have a cost smaller than the threshold; then the path 
starting in $s$ and ending on $t$ has cost less than $0$ 
and is exactly the boundary wall of the percolation process on the 
lattice with weights $\omega(\mathcal{G})$ and threshold $0.5$. 
Then the measure of the optimal paths $p_{s,e}$ is the same as that of  
critical site percolation exploration path on the honeycomb lattice, 
that is (see \cite{Camia}), in the termodynamic limit  (infinitesimal 
lattice spacing), SLE measure with $\kappa =6.$

\subsection{Motivations}
In the study of domain walls in disordered systems it happens that the 
union of 
the domain walls is a tree; this is the case for example in Ising Spin 
Glasses for domain walls constrained to start from a fixed point, it is the 
case also for the boundary given by the symmetric difference of opportune 
matchings on planar graph \cite{FicheraSportiello}. 

Such ubiquity suggests to search for a disordered model not only such as 
to reproduce the SLE distribution of probability, but also such 
that the union of the optimal paths was a tree and, both for his 
mathematical properties and for numerical reasons, we wanted this tree 
to be easy to find.

One of the properties of MST that make it easy to find (in computational 
sense) is a locality property (see \ref{appendix:MST}) similar to the 
locality property of SLE for $\kappa = 6$.

\section{Results}
\label{sec:Results}

\subsection{The Samples}
We simulated numerically rectangular samples of sizes ranging 
from $32\times 32$ to $1024\times 1024$. 

For clarity, suppose that the four 
vertices of the rectangle are $(0,0)$, $(0,a)$, $(b,0)$ and $(a,b)$. Let 
$s = (a/2 , 0)$ be the point at one half of the bottom edge of the rectangle 
and $e = (a/2 , b)$ be the point at one half of the top edge. We will consider 
four different boundary conditions and a slightly different model:

\begin{itemize}
\item $Free$: the weights on the boundary plaquettes are extracted as 
the bulk ones.
\item $SLE-like$: the weights on the boundary plaquettes on the left of 
$s$ and $t$ are constrained to have a weight higher than 
$\theta$ and the other boundary plaquettes are constrained to have a weight 
lower than $\theta$. 
\item $SLE/Free$: The boundary plaquettes on the right edge of the rectangle 
are constrained to be higher than $\theta$ and the ones on the left edge are 
constrained to have a weight lower than $\theta$. The weights on the top edge 
and on the bottom edge are unconstrained (free).  
\item $Repulsive$: All the boundary plaquettes are constrained to have a 
weight higher than the threshold $\theta$.
\item $Random$: the weights on the edges are i.i.d. variables. Remark that it 
is not a peculiar choice of the boundary conditions for our model. It is 
another model: the random MST model. We study this random measure on the 
weights of the edges just for comparison with this well known model.
\end{itemize}

In order to study systems at criticality we mainly concentrate on a   
$\theta$ equal to the percolation threshold for site percolation (0.5 on 
the honeycomb lattice, 0.5927463 on the square one). On the square models 
some boundary 
conditions ($SLE-like$, $SLE-free$) break the left-right symmetry because 
the critical threshold is different from $0.5$, so we simulated our model 
on the square lattice also with $\theta = 0.5$, we expected a trivial limit 
for these paths, the fact that we did not observe it means that the scaling 
limit was not reached by the numerical simulations.

\subsection{Observables}
\label{sec:Observables}
We measured the fractal-dimension of the paths. We know that the 
fractal-dimension of SLE-Walks is linked to the parameter $\kappa$ by 
the relation $d = 1 + \kappa /8 $. 

We measured left-passage probability, which because of dilation 
invariance, has to be  a function only of the radial coordinate 
on the half plane. Schramm's formula (see~\cite{Schramm}) links the shape 
of the 
left-passage probability to the parameter $\kappa$. As we observed (see 
\cite{FicheraSportiello}) it can happen, for disordered systems, 
that the parameter $\kappa$ found by left-passage probability and the 
fractal dimension are not compatible, indicating that SLE and minimizing paths 
are not equal in measure.

As it is usual in literature (\cite{Hartmann-Amoruso} 
\cite{BernardLeDoussal}) we consider both the path starting in 
$s$ and ending in $t$ and the optimal path among 
the ones starting on the bottom and ending on the top. Notice that, for  
$SLE-like$ boundary conditions, the two paths coincide: the optimal path 
connecting the bottom to the top is also the one starting in $s$ and 
ending in $t$. 

If we want to compare the left-passage probability of paths on rectangles 
with Schramm's formula we have to transform the domain into the half 
plane. For the path starting in $s$ and ending in $t$ we choose the 
conformal transformation that 
maps $(a/2 , 0)$ in $(0,0)$, $(a/2,b)$ in $\infty$, $(0,0)$ in $(-1,0)$ and 
$(a,0)$ in $(1,0)$. For the path connecting the top to the bottom we consider 
the conformal transformation that maps the rectangle to the semi-annulus 
such that the vertices of the rectangle are sent on the vertices of the 
rect angles of 
the half annulus and $(0,b/2)$, $(a,b/2)$ are sent respectively in $(-1,0)$ 
$(1,0)$. This transformation sends the rectangle to the half plane only in 
the limit $b/a \rightarrow \infty$, the limit considered  in 
$\cite{BernardLeDoussal}$ to study the horizontal displacement. For $b/a 
< \infty$ boundary effects are observed at top and bottom.

The horizontal displacement is the difference $\Delta(x)$ between the 
abscissae of starting and ending point for the optimal path connecting the 
top of the square to the bottom. We measured the average value 
of ${\Delta x}^2$, with $\Delta x$ expressed in unit of $a$, the horizontal 
lenght of the rectangle, so that $\Delta x \in [-1,1]$ for every path.

\subsubsection{Fractal Dimension}
The fractal dimensions of the curves is measured by comparing the 
number of steps of the paths in lattices of different sizes. Having  
fixed the boundary conditions, the fractal dimension is independent of 
the path considered.
\begin{center}
\begin{tabular}{|r||c|c|}
\hline
 & Square & Honeycomb \\
\hline 
\hline 
$Free$      & 1.21 $\pm$ 0.01 & 1.75 $\pm$ 0.01 \\
$SLE-like$  & 1.20 $\pm$ 0.01 & 1.75 $\pm$ 0.01 \\
$SLE/Free$  & 1.22 $\pm$ 0.01 & 1.75 $\pm$ 0.01 \\
$Repulsive$ & 1.22 $\pm$ 0.01 & 1.75 $\pm$ 0.01 \\
$Random$    & 1.21 $\pm$ 0.01 & 1.22 $\pm$ 0.01 \\
\hline
\end{tabular}
\end{center}

\subsubsection{Left-Passage Probability}
Left passage probabilities (the probability for a point in the domain to 
be at the left or at the right of the path) have been measured in 
rectangular domains. For the path with ends in $s$ and $t$ we 
transformed the domain to the half plane to compare the measured 
probability (over $10^5$ samples) to the Schramm formula: 
$$
1/2 + \frac{\Gamma(\frac{4}{\kappa})}{\sqrt{\pi} \Gamma
(\frac{8-\kappa}{2\kappa})} \tan t \cdot \phantom{}_2 F_1 
\left[ \frac{1}{2} , \frac{4}{\kappa}, \frac{3}{2}, -\tan^2 (t)\right] 
$$
where $t$ is the angle subtended between the ray in $s$ and the real axis.
 
For the paths with free ends we compared measured probabilities wih the 
formula via the identification of $x$ (the coordinate on the rectangle) with 
the angle $\theta$ on the half plane. 
\begin{center}
\begin{tabular}{|r||c|c|}
\hline
Path $s \rightarrow t$ & Square & Honeycomb  \\
\hline 

\hline 
$Free$      &  2.8 $\pm$ 0.1 &  2.7 $\pm$ 0.1  \\
$SLE-like$  &  XXX           &  6.0 $\pm$ 0.1  \\
$SLE/Free$  &  XXX           &  XXX            \\
$Repulsive$ &  XXX           &  XXX            \\
$Random$    &  2.8 $\pm$ 0.1 &  2.8 $\pm$ 0.1  \\
\hline
\hline
Optimal path & Square & Honeycomb  \\
\hline
 
\hline 
$Free$      &  3.2 $\pm$ 0.1 & 2.9 $\pm$ 0.1  \\
$SLE/Free$  &  XXX           & XXX            \\
$Repulsive$ &  5.9 $\pm$ 0.1 & XXX            \\
$Random$    &  3.2 $\pm$ 0.1 & 3.2 $\pm$ 0.1  \\
\hline
\end{tabular}
\end{center}
The entries marked with XXX correspond to measured left passage probabilities 
that do not fit with Schramm's formula for any value of $\kappa$.

Notice that the critical model (on both the lattices) has compatible 
values of $\kappa$ for $Free$ and $Random$ boundary 
conditions, but they are very different for $Repulsive$ boundary conditions. 
\subsubsection{Horizontal Displacement}
We observe that when the height of the rectangle is bigger than the 
base, the position of the starting point and the position of the 
ending point are uncorrelated. As a consequence the average value 
of ${\Delta x}^2$ 
is constant for $b \gg a$ and converges to a value $\langle {\Delta x}^2 
\rangle$. For 
$b \ll a$ we measure the exponent $l$ in $\Delta x (b/a) \sim (b/a)^l$ 
\\ 
\begin{center}
\begin{tabular}{|r||c|c|}
\hline
Square & $\langle {\Delta x}^2 \rangle$ & l \\
\hline 

\hline 
$Free$      & 0.134 $\pm$ 0.05  &  2.07 $\pm$ 0.03 \\
$SLE/Free$  & 0.126 $\pm$ 0.002 &  2.10 $\pm$ 0.16 \\
$Repulsive$ & 0.24  $\pm$ 0.01  &  2.13 $\pm$ 0.09 \\
$Random$    & 0.128 $\pm$ 0.005 &  2.14 $\pm$ 0.26 \\
\hline
\hline
Honeycomb & $\langle {\Delta x}^2 \rangle$ & l \\
\hline 

\hline 
$Free$      &  0.104 $\pm$ 0.001 &  2.22 $\pm$ 0.14 \\
$SLE/Free$  &  0.190 $\pm$ 0.01  &  2.05 $\pm$ 0.02 \\
$Repulsive$ &  0.190 $\pm$ 0.01  &  2.05 $\pm$ 0.12 \\
$Random$    &  0.129 $\pm$ 0.005 &  2.24 $\pm$ 0.19 \\
\hline
\end{tabular}
\end{center}

\subsubsection{Conformal Invariance of the Trees}
We have investigated the conformal invariance of the Minimum Spanning Tree. 
As we know \cite{Wilson}, for the Random Spanning Tree the conformal 
invariance does not hold. To test conformal invariance we measure 
the distribution of probability of the triple point $T$ on the square. 
The triple point is defined as the unique site in the tree connected 
to $(0,0)$, $(0,b)$, $(a,0)$ by 
three paths with null intersection. 
We transform conformally the rectangle into a disk so to map the 
points  $(0,0)$, $(0,b)$, $(a,0)$ on the vertices of an equilateral triangle 
inscribed in the disk. If conformal invariance for the tree holds 
(as it happens for instance for the uniform spanning trees), the 
transformed distribution of probability should be invariant under rotations 
of $2\pi /3$ of the disk.
This test has been done for all the models with boundary conditions that do 
not break conformal invariance ($Free$ and $Repulsive$) and has shown that 
conformal invariance does not hold for the trees we defined.


\section{Conclusions and Perspectives}
\label{sec:conclusions}
Several surprising facts emerged from the numerical simulations. The fractal 
dimension of paths is irrespective of the boundary conditions, but it 
depends dramatically on the kind of lattice. The left-passage probability, 
also on the honeycomb lattice, is not compatible with the fractal dimension 
for every choice of the boundary conditions different from the standard one, 
anyway on the square lattice with $Repulsive$ boundary conditions and 
at the critical percolation threshold, the left-passage probability 
obtained is consistent with $\kappa = 6$. These 
facts are not well understood and need more investigations also on 
different lattices.

Given two vertices $i$, $j$, we say that they are connected if the 
cost of the minimal path between $i$ and $j$ is less than $0$.
The behaviour of the connection probability could be studied both 
numerically and theoretically using CFT's tools, as in \cite{Ziff} for 
critical percolation.
The structure of the connected domains is better understood in the 
scheme of the Krushkal's algorithm (see \ref{appendix:MST}). 
The SLE boundary conditions are peculiar because all the boundary sites 
but two are disconnected.  

It is possible to define a process of growth of trees in the scheme 
of Prim's algorithm, in fact one could start to grow the tree from a 
starting point on the boundary and progressively increasing 
it with Prim's algorithm. This is the definition of a process 
of growth for Spanning 
Trees. It  would be interesting to understand if, using  the 
reparametrization of the time such that the rate of increase of the 
capacity be constant, the continuum limit of this evolution process 
makes sense.
Notice that $\mathrm{SLE}_6$ is recovered as the growth of the tree with
opportune boundary conditions.

This method to define growth processes for trees such that, with opportune 
boundary conditions, $\mathrm{SLE}_6$ is recovered could be easily generalized to 
other spin models. 
In fact, given a spin configuration extracted with the Gibbs measure and the 
usual boundary conditions to force a boundary wall to exist starting on $s$ 
and ending on $t$, we 
need only to associate a weight bigger than $\theta$ to sites with up spins 
and smaller than $\theta$ to sites with down spin. Then, the minimum spanning 
tree on the honeycomb lattice with weights induced by equation 
(\ref{eqn:makeweights}) will contain by construction the boundary 
between up spins and down spins starting in $s$ and ending in $t$.

In this draft we studied only the optimal spanning tree; it could be 
interesting also to study the low temperature behaviour: the almost 
optimal trees.

One could  investigate the stability of walks under perturbations 
of the in\-stance. It is not a very hard task when working in the scheme of 
Krushkal's algorithm, thanks to MST properties.


\appendix

\section{Minimum Spanning Tree}
\label{appendix:MST}
Let $\mathcal{G} = (V,E)$ be a connected graph with $N$ vertices, a spanning tree 
$T$ is a loopless subgraph with $N-1$ edges. i.e. it is a tree (loopless 
and connected) and it is spanning (every vertex in $V$ has at least one 
incident 
edge in $T$).
Given a weighted graph $(\mathcal{G} , W)$ with real weights on the 
edges, a Minimun Spaning Tree is a Spanning Tree of minimum weight 
(where the weight of a tree $T$ is the 
sum of the weights of the edges in $T$).

MST have an important property that we would like to stress: given a 
subset $V^\prime$ of the vertices $V$ of $\mathcal{G}$, let $B_{V^\prime}$ be the edges 
on the boundary beetween $V^\prime$ and its complement $\bar{V^\prime}$ 
then the MST contains the edge $e$ of minimum weight among the edges in $B_V$.

This property allows some local algorithms to work. By locality we mean 
that we do not need to know all the weights over the whole graph to find 
one edge in $B_{V^\prime}$ that will be also in $T$.
Notice also that if the MST on $(\mathcal{G}(V,E),W)$ restricted to a subset $V^\prime$ 
of $V$ is connected then it coincides with the MST of $(\mathcal{G}(V^\prime 
,E_{V^\prime}),W)$.

There are two basics strategies for the search of the MST, 
one consists on the progressive increase of a minimum tree until it is 
spanning, the other one consists on the progressive coalescence of the trees 
in a spanning forest (collection of trees) until a single tree is obtained. 
 
In the first one (Prim's algorithm) one starts with a given vertex 
$i \in V^\prime$ and 
progressively increases this set with the minimal edge in $B_{V^\prime}$ 
until every vertex is in $V^\prime$. 
In the other one (Krushkal's algorithm) one starts with a forest consisting of 
all the vertices and no edges, then one starts to increment the set of edges in 
the MST with the edge of minimal weight, and goes on adding minimal edges 
unless a cycle would result.

In our numerical simulations we used the Kruskal algorithm which is 
pol\-y\-no\-mi\-al ($|E|\ln |E|$) in the number $|E|$ of edges of the graph.
 
We remark that it is easy to write polynomial algorithms to study the 
excited states of the MST (spanning trees with almost minimum cost), 
so to investigate low temperature properties of the model.


                   \acknowledgments

I thank P.~Contucci and R.~Santachiara for encouragments, A.~Bedini 
M.~Ghe\-rar\-di and A.~Nigro for their several suggestions and A.~Sportiello 
for his essential teachings. 



\begin{thebibliography}{100}



\bibitem{BauerBernard}
M.~Bauer and D.~Bernard,
{\it 2D growth processes: SLE and Loewner chains}
Phys.~Rept (\textbf{432}) 115-221 (2006) ({\tt math-ph/0602049}).

\bibitem{Hartmann-Amoruso}
C.~Amoruso, A.K.~Hartmann,M.B.~Hasting and M.A.~Moore 
{\it Conformal Invariance and SLE in Two-Dimensional Ising Spin Glasses} 
PRL {\textbf{97}}, 267202 (2006)


\bibitem{BernardLeDoussal}
D.~Bernard,P.~Le~Doussal, A.A.~Middleton
{\it Are Domain Walls in 2D Spin Glasses described by Stochastic 
Loewner Evolutions?}
PRB {\textbf{76}} 020403(R) (2007) ({\tt cond-mat/0611433})

\bibitem{FicheraSportiello}
D.F., A.~Sportiello,
{\it Double Assignment model and SLE },
in prep.

\bibitem{Wilson}
D.B.~Wilson
{\it On the Red-Green-Blue Model},
PRE {\textbf{69}} 037107 (2004) ({\tt cond-mat/0212042})

\bibitem{Ziff}
P.~Kleban, J.J.H.~Simons, R.M.~Ziff 
{\it Anchored critical percolation cluster and 2D electrostatic}
PRL {\textbf{97}}, 115702 (2006)

\bibitem{Camia}
F.~Camia, C.M.~Newman
{\it Critical Percolation Exploration Path and $\mathrm{SLE}_6$: 
a proof of convergence}
{\tt math/0604487}

\bibitem{Schramm}
O.~Schramm
{\it A Percolation Formula}
Electronic Comm. Probab. {\textbf 8}, paper 12, 2001 ({\tt math.PR/0107096})


\end{thebibliography}
\end{document}